\documentclass[aps,prb,twocolumn,groupedaddress,floatfix,showpacs,amsmath,amssymb]{revtex4}   
\usepackage{graphicx}
\usepackage{amsmath}
\usepackage{psfrag}
\begin{document}
\newcommand\ket[1]{|#1\rangle}
\newcommand\bra[1]{\langle#1|}
\newcommand\braket[2]{\left\langle#1\left|#2\right.\right\rangle}

\title{Electron counting with a two-particle emitter}

\author{Janine Splettstoesser$^{1}$,
Sveta Ol'khovskaya$^{2}$,
Michael Moskalets$^{1,2}$,
Markus B\"uttiker$^{1}$
}
\affiliation{$^{1}$D\'epartement de Physique Th\'eorique, Universit\'e de
  Gen\`eve,  CH-1211 Gen\`eve 4, Switzerland\\
$^{2}$Department of Metal and Semic. Physics, NTU ``Kharkiv Polytechnic
Institute'', 61002 Kharkiv, Ukraine
}

\date{\today}

\begin{abstract}
We consider two driven cavities (capacitors) connected in series via an edge
state. The cavities are driven such that they emit an electron and a hole in
each cycle. Depending on the phase lag the second cavity can effectively
absorb the carriers emitted by the first cavity and nullify the total current
or the set-up can be made to work as a two-particle emitter.
We examine the precision with which the current can be nullified and with which
the second cavity effectively counts the particles emitted by the first
one. To achieve single-particle detection we examine pulsed cavities.  
\end{abstract}
\pacs{72.10.-d,73.23.-b}
% 72.10.-d Theory of electronic transport; scattering mechanisms
% 73.23.-b Electronic transport in mesoscopic systems
\maketitle
%%%%%%%%%%%%%%%%%%%%%%%%%%%%%%%%%%%%%%%%%%%%%%%%%%
\section{Introduction}
%%%%%%%%%%%%%%%%%%%%%%%%%%%%%%%%%%%%%%%%%%%%%%%%%%
The dynamics of a quantum coherent capacitor connected via a single contact
to an electron reservoir have attracted experimental and theoretical
interest. A capacitor connected via a quantum point contact (QPC) to an edge
state shows mesoscopic capacitance oscillations and a quantized charge
relaxation resistance~\cite{buttiker93,gabelli06,nigg06,wang07,nigg08}. In addition a recent
experiment demonstrated an ``electron gun'' emitting and absorbing a single
electron in every oscillation cycle~\cite{feve07}. The emission
process~\cite{moska08,keeling08,olkho08} injects an electron into states above the Fermi level,  
whereas absorption of an electron leaves a hole below the Fermi energy. 
The invention of Lasers revolutionized optics. Similarly, single electron
injectors either using capacitors or quantized electron pumps~\cite{blumenthal07,kaestner08,fujiwara08,maire08,kaestner08_2}
provide novel, coherent sources for electronics. \par
\begin{figure}
\includegraphics[width=3.in]{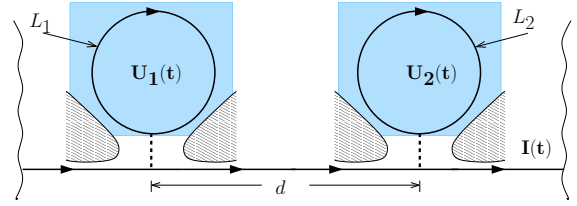}
\caption{
  (Color online) Two driven cavities (mesoscopic capacitors), formed
  with quantum point contacts, are coupled in series by an edge state. 
  Time-dependent potentials $U_1(t)$ and $U_2(t)$ act 
  homogeneously on the regions of the two cavities. 
  }
\label{fig_model}
\end{figure}

It is a challenging task to detect the electrons with the speed they
were emitted with. In modern experiments 
the dynamics of single electron transport through a
mesoscopic system is often explored experimentally using as a charge
detector, either a radio-frequency single-electron
transistor~\cite{lu2003,fujisawa2004,bylander2005} or a
QPC~\cite{gustavsson06,fujisawa06,fricke07,reilly}.  
However the speed of these detectors is not sufficient to detect
electrons with a nanosecond resolution. 
To circumvent this problem, we propose 
as a fast detector a device which is analogous to the emitter: a quantum capacitor, such as used in~\cite{feve07,gabelli06}. 
Such a detector is able to register particles as fast as
an emitter can inject them into the quantum circuit. 
We therefore consider a system consisting of two quantum cavities coupled in
series by a single edge state and modulated by in general
different, periodically-varying potentials, both with frequency $\Omega$, as
shown in Fig.~\ref{fig_model}. 
The charge
emitted by the first cavity is detected by nullifying the \textit{total}
current with the use of the modulation of the second cavity. Namely, the
potential $U_2(t)$ can be chosen in such a way that the total current
vanishes. In general the current $I(t)$ consists of a series of pulses 
corresponding to electrons and holes emitted by either of the cavities.
However if the time when an electron was emitted by the first cavity coincides
with the time when a hole was emitted by the second cavity the total current
$I(t)$ is suppressed. This electron-hole annihilation process can be viewed as 
the reabsorption by the second cavity of an electron emitted by the first
cavity and it can be used to count electrons. If the counting efficiency is
perfect the total current vanishes completely~\cite{comment_random}.\par 
Since the capacitor system generates an AC current it is convenient to
investigate the degree of the current suppression by studying the square of
the current
%~\cite{notenoise}
 integrated over one period $2\pi/\Omega$,  
\begin{equation} 
\label{eq_scip}
\langle I^{2}\rangle = \int_{0}^{2\pi/\Omega} dt\, I^2(t)\,.
\end{equation}
Note that $\langle I(t)^2\rangle$ is different from the noise, see Ref. \onlinecite{moska08,keeling08,olkho08}. 
We develop the conditions for nullifying the total current and investigate the measuring precision.
Alternatively, being driven in phase such a double-capacitor system can serve
as a two-electron (two-hole) emitter.  
%%%%%%%%%%%%%%%%%%%%%%%%%%%%%%%%%%%%%%%%%%%%%%%%%%
\section{Model and Formalism}
%%%%%%%%%%%%%%%%%%%%%%%%%%%%%%%%%%%%%%%%%%%%%%%%
The system consists of two cavities with edge states of circumference $L_1$
and $L_2$ connected via QPCs with the reflection (transmission) 
amplitudes $r_1$ ($t_1$) and $r_2$ ($t_2$) to an edge state of length  $d$
and modulated by time-dependent potentials $U_1(t)$ and $U_2(t)$ respectively.
A particle with energy $E$ entering the cavity $j$ picks up a phase 
$kL_j$, which is the kinetic phase of the guiding center
motion~\cite{fertig87}. The time $\tau_j$ that a
particle spends for one revolution in the cavity $j$ is related to the
cavity's level spacing, $\Delta_j=h/\tau_j$. Due to the time-dependent
potential $U_j(t)$ an additional time-dependent  phase
$\Phi_j^q(t)=\frac{e}{\hbar}\int_{t-q\tau_j}^t dt'U_j(t')$ is accumulated in
the cavity during $q$ revolutions. 
The separate cavities can be
described by a time-dependent scattering matrix for a particle with incoming
energy $E$, leaving the system at time $t$, given by the Fabry-Perot like
expression 
\begin{eqnarray}\label{eq_single_smatrix}
S_j(E,t) & = & r_j+t_j^2\sum_{q=1}^\infty
r_j^{q-1}e^{iq
kL_j-i\Phi_j^q(t)}.
\end{eqnarray}
With the time $\tau_d$ which the particle spends in the connecting edge state,
the scattering matrix of the full system is  
\begin{eqnarray}\label{eq_full_s}
&& S_\mathrm{tot}(t,E)  =  
\sum_{p,q=0}^{\infty}\left(\left(r_2^*\right)^{-1}\delta_{p0}
+t_2^2 r_2^{p-1}e^{ip
kL_2-i\Phi^p_2(t)}
\right)\nonumber\\
&&\cdot\left(\left(r_1^*\right)^{-1}\delta_{q0}
+t_1^2 r_1^{q-1}e^{iq
kL_1-i\Phi^q_1(t-\tau_d-p\tau_2)}
\right)e^{i
kd}\ .
\end{eqnarray}
A Floquet scattering matrix approach~\cite{moska08,keeling08,olkho08}, 
used to deal with quantum pumping~\cite{brouwer98,avron00,vavilov05,arrachea05,splett05,graf08},
enables us to investigate
the dynamics of the system beyond the linear-response regime and adiabatic
approximations. The full 
time-dependent current response to a periodic
modulation with frequency $\Omega$ is
\begin{eqnarray}\label{eq_current_general}
I(t)& = & \frac{e}{h}\int
dE\sum_{n=-\infty}^{\infty}\left[f(E) - f(E+\hbar n\Omega)\right]\\
&& 
\cdot\frac{\Omega}{2\pi}\int_0^{2\pi/\Omega} dt' e^{in\Omega(t-t')} 
S_\mathrm{tot}^*(t',E)S_\mathrm{tot}(t,E)\ .\nonumber
\end{eqnarray}
In the following we analyze the conditions to achieve efficient particle counting in the
double-capacitor system by nullifying the total current and discuss the
precision. 
%%%%%%%%%%%%%%%%%%%%%%%%%%%%%%%%%%%%%%%%%%%%%%%%%%
\section{Results}
%%%%%%%%%%%%%%%%%%%%%%%%%%%%%%%%%%%%%%%%%%%%%%%%%%
Inserting the total scattering matrix, given in Eq. (\ref{eq_full_s}) into the current formula of Eq. (\ref{eq_current_general}), we obtain a general result for the total current due to a harmonic modulation of the system. We first investigate the adiabatic regime, specifying results at zero and at high temperatures. Subsequently corrections to the adiabatic results and the strongly nonadiabatic limit are considered.
%%%%%%%%%%%%%%%%%%%%%%%%%%%%%%%%%%%%%%%%%%%%%%%%%%
\subsection{Adiabatic response}
%%%%%%%%%%%%%%%%%%%%%%%%%%%%%%%%%%%%%%%%%%%%%%%%%%
In the following we calculate the current response to two 
potentials $U_j(t)=\bar{U}_j+\delta U_j(t)$. 
In the adiabatic limit, $\Omega\rightarrow 0$, where the time scale set by the
modulation is much larger than the time particles spend in the cavities and the connecting edge state, 
we expand Eq.~(\ref{eq_current_general}) in first order $\Omega$. The current
$I^{(1)}(t)$ is related to the instantaneous densities of states 
  $\nu_j=\nu_j(t,E)=\frac{1}{2\pi  i}S_j^*(E-eU_j(t))\frac{\partial
  S_j(E-eU_j(t))}{\partial E}$ of the two cavities
\begin{eqnarray}\label{eq_lowest_order}
I^{(1)}(t)  =  
e^2\int dE\left(-f'(E)\right)
\left[\nu_1\frac{\partial U_1(t)}{\partial
  t}+\nu_2\frac{\partial U_2(t)}{\partial t}\right].
\end{eqnarray}
With the transmission $T_j=\left|t_j\right|^2$ the density of states is
\begin{equation}\label{eq_DOS}
 \nu_j(t,E)=\frac{1}{\Delta_j}\frac{T_j}{2-T_j-2\sqrt{1-T_j}\cos\phi_j(E,t)}.
\end{equation}
The phase
$\phi_j(E,t)$ can be written as the sum of a time-dependent and a
time-independent contribution 
\begin{eqnarray}\label{eq_energy_phase}
\phi_j(E,t) =: -2\pi e\delta
  U_j(t)/\Delta_j+2\pi \chi_j(E),
\end{eqnarray} 
with $2\pi \chi_j(E) = \frac{\tau_j}{\hbar}\left(E-\mu\right) +
k(\mu)L_j- 2\pi e\bar{U}_j/\Delta_j+\phi^r_j .$ The phase of the respective QPC's 
reflection coefficient is given by $\phi^r_j$, with $r_j=|r_j|exp(i\phi^r_j)$. 
The detuning $\chi_j\Delta_{j}$ defines the position of the quantum level in
the cavity $j$ with respect to the Fermi level $\mu$ at zero driving
amplitude, $\delta U_j(t) = 0$. 
Thus to lowest order in frequency the current consists of a sum of separate
contributions of the two cavities. This means that the total time-dependent
current can be 
nullified, whenever the phase difference $\delta_2-\delta_1$ of the two
harmonic modulations
is equal to $\pi$ and when the amplitudes together with the cavity parameters are adjusted in an appropriate way.
We now choose a harmonic time-dependence, 
$\delta U_j(t)=U_j\cos(\Omega t+\delta_j)$, with $U_j>0$.   \par
\begin{figure}[t]
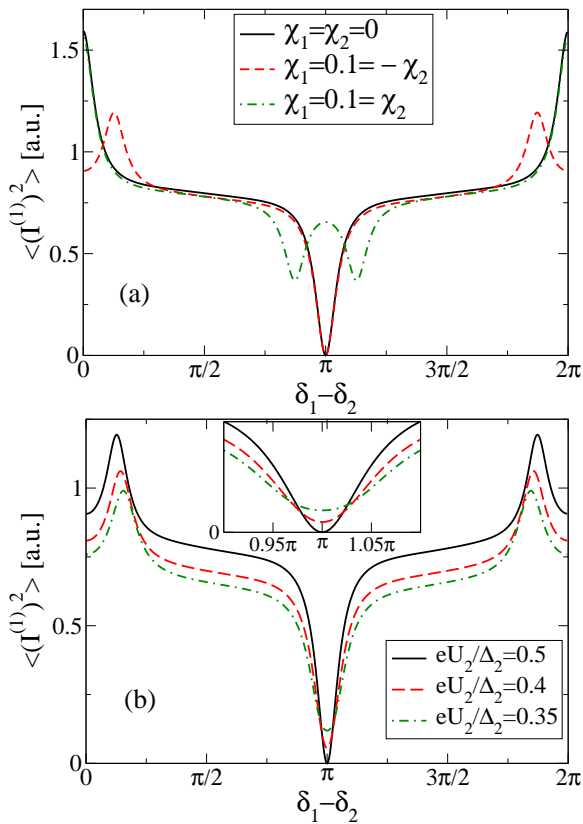

\includegraphics[width=3.0in]{fig2a.eps}
\includegraphics[width=3.0in]{fig2b.eps}
\caption{(Color online) Zero-temperature limit of the averaged square current
  $\langle (I^{(1)})^2\rangle$ from Eq.~(\ref{eq_lowest_order}) as a function
  of the phase difference of the modulation potentials. The ratio between
  modulation amplitude and the level spacing of the first cavity is given by
  $eU_1/\Delta_1=0.5$. The transmission probabilities of the QPCs of the two
  cavities are $T_1=T_2=0.4$. (a) Different detuning and
  $eU_2/\Delta_2=0.5=eU_1/\Delta_1$. (b) Fixed detuning $\chi_1=0.1=-\chi_2$
  and different values of $eU_2/\Delta_2$.}
\label{fig_current_average}
\end{figure}
%%%%%%%%%%%%%%%%%%%%%%%%%%%%%%%%%%%%%%%%%%%%%%%%%%
 \subsection{Current nullification at $k_\mathrm{B}T=0$}
%%%%%%%%%%%%%%%%%%%%%%%%%%%%%%%%%%%%%%%%%%%%%%%%%%
In Fig.~\ref{fig_current_average} we plot the time integral of the
squared 
current,
 Eq.~(\ref{eq_lowest_order}), 
as a function of the phase difference of the potentials
 given
for different
choices  of detuning $\chi_1\Delta_1$,
$\chi_2\Delta_2$ and for different
values for $eU_2/\Delta_2$. 
To understand these plots, we consider the limit of very low transmission at
the QPC's, $T_j\ll 1$. Then the instantaneous density of states,
Eq.~(\ref{eq_DOS}), of the cavities takes the form of a sum of Breit-Wigner
resonances, around the zeroes of the 
quantity $\phi_j(E,t)$, defined in
Eq.~(\ref{eq_energy_phase}), to be taken mod$2\pi$.
We 
take
the modulation amplitude $U_j$ 
to be smaller than
half the level spacing $\Delta_j$ and larger than the detuning
$\chi_j\Delta_j$, such that one electron and one hole are emitted per cycle. We
consider particles with energies equal to the Fermi energy. \par
When periodically driving the potentials $U_j(t)$,
the densities of
states have 
a peak at the Fermi energy around resonance times $t_j^+$ and
$t_j^-$. 
To lowest order $\Omega$, 
the 
current pulse
generated,
Eq.~\ref{eq_lowest_order}, is expressed in terms of the resonance times,
$t_j^{+/-}$, and the half-widths of the pulses, $w_j$, 
\begin{subequations} 
\begin{eqnarray}\label{eq_res_time}
\Omega\ t_j^{+/-} & = & - \delta_j
\pm\arccos\left(\frac{\chi_j\Delta_j}{eU_j}\right)\\ 
\Omega\ w_j & = &\label{eq_halfwidth}
\frac{1}{2\pi}\frac{T_j\Delta_j}
{2eU_j} \left[1-\left(\frac{\chi_j\Delta_j}{eU_j}\right)^2\right]^{-1/2}\ .
\end{eqnarray}
\end{subequations}
We are interested in a
situation where during the driving process an electron and a hole are fully
emitted, separately from each other, and therefore the distance between the
resonance times $t_j^\pm$ is much larger than the 
width of the current pulse, $|t_j^+-t_j^-|\gg w_j$. We find
\begin{eqnarray}\label{eq_breit_wigner}
&&\langle(I^{(1)})^2\rangle =
\frac{e^2}{\pi}\left[\frac{1}{w_1}+\frac{1}{w_2}\right]+
\frac{2e^2}{\pi}\left[
L(t_1^+-t_2^+)
\right.\nonumber\\
&&\left.
+L(t_1^--t_2^-)-L(t_1^+-t_2^-)-L(t_1^--t_2^+)
\right]\ .
\end{eqnarray}
where we introduce the Lorentzian
$L(X)=\left(w_1+w_2\right)/\left[X^2+(w_1+w_2)^2\right]$. Its
arguments $\left(t_1^\pm-t_2^\pm\right)$  
are taken mod$2\pi$, 
in the interval $[-\pi/\Omega,\pi/\Omega]$.
The four Lorentzians contribute only if the respective resonance times are
close to each other compared to the width of the current pulse.
If the first two Lorentzians contribute, 
two particles are emitted by the system at the same time, either two electrons
or two holes and 
$\langle(I^{(1)})^2\rangle$ is maximized.  
We are instead interested in the situation where both of the last
two terms contribute, meaning that one cavity emits a hole approximately at
the same time as the other emits an electron and vice versa.
The conditions for
nullifying the current exactly are 
\begin{subequations} \label{eq_conditions}
\begin{eqnarray}
\delta_1-\delta_2 & = & \pi\\
 \chi_1/T_1 & = & -\chi_2/T_2\\
eU_1/(T_1\Delta_1)  & = & eU_2/(T_2\Delta_2)\ ,
\end{eqnarray}
\end{subequations} 
Experimentally these conditions can be obtained by tuning the phase $\chi_j$,
the amplitude of the time dependent part 
and the phase difference of the potentials. Close to these conditions,
$\langle(I^{(1)})^2\rangle$ as a function of the phase difference has a 
pronounced dip 
\begin{equation}
\langle(I^{(1)})^2\rangle
=\frac{2e^2}{w\pi}\frac{\left(\delta_1-\delta_2-\pi\right)^2}
{\left(\delta_1-\delta_2-\pi\right)^2+4w^2\Omega^2}\
,  
\end{equation}
where $\left(\delta_1-\delta_2-\pi\right)$ is
taken mod$2\pi$ on the
interval $\left[-\pi,\pi\right]$. 
In Fig.~\ref{fig_current_average} (a) we show $\langle(I^{(1)})^2\rangle$
as a function of
$\delta_1-\delta_2$  
at finite transmission probability of the QPCs for
$\frac{eU_1}{T_1\Delta_1}=\frac{eU_2}{T_2\Delta_2}$. 
Whenever the maximum is at $\delta_1-\delta_2=0$, both the two
electrons and the two holes are respectively emitted at the same time (solid
and dashed-dotted curve). Whenever the minimum is at $\delta_1-\delta_2=\pi$,
any emitted electron 
is annihilated by a hole at the same time (solid and dashed line).
If the current pulses of
an electron of one cavity and a hole of the other are both
coinciding but the width of the pulses are different, the distance of the 
minimum of $\langle(I^{(1)})^2\rangle$ from zero is 
\begin{eqnarray}
\langle(I^{(1)})^2\rangle =
\frac{e^2}{\pi}\frac{\left(w_1-w_2\right)^2}
{w_1w_2\left(w_1+w_2\right)}\ ,
\end{eqnarray}
showing a smooth dependence on the system parameters. It guarantees the
robustness of the dip against small deviations from the ideal conditions. This
minimum for the more general case of Eq.~(\ref{eq_lowest_order}) is shown 
in Fig.~\ref{fig_current_average} (b). 
%%%%%%%%%%%%%%%%%%%%%%%%%%%%%%%%%%%%%%%%%%%%%%%%%%
\subsection{High temperatures, $k_\mathrm{B}T\gg\Delta_j$}
%%%%%%%%%%%%%%%%%%%%%%%%%%%%%%%%%%%%%%%%%%%%%%%%%
In this regime the quantized emission is destroyed. 
However the current nullification can still be achieved and, e.g., be used to
tune the parameters of the cavities. At high temperatures we use  $\nu_{j} = 1/\Delta_{j}$ in Eq.~(\ref{eq_lowest_order}). Then from Eq.\,(\ref{eq_scip}) we find that
the time integral of the square of the low-frequency current
takes a particularly simple form
\begin{eqnarray}\label{eq_highT_first}
\frac{\langle (I^{(1)})^2\rangle}{\pi e^2\Omega} =  
e^2\left(\frac{U_1^2}{\Delta_1^2}+2\frac{U_1
  U_2}{\Delta_1\Delta_2}\cos(\delta_1-\delta_2)+\frac{U_2^2}{\Delta_2^2}\right)
.  
\end{eqnarray}
It shows a cosine-like behavior as a function of the
phase-difference, in contrast to the zero-temperature result, where the width
of the dips (peaks) is  determined by $w_j$. Independently of the detuning of
the two cavities and the transmission of the QPCs, $\langle
(I^{(1)})^2\rangle$ is exactly zero when 
$eU_1/\Delta_1=eU_2/\Delta_2$ and $\delta_1-\delta_2=\pi$ and deviates from
zero at $\delta_1-\delta_2=\pi$, by 
$\pi e^2\Omega\left(eU_1/\Delta_1-eU_2/\Delta_2\right)^2$.
%%%%%%%%%%%%%%%%%%%%%%%%%%%%%%%%%%%%%%%%%%%%%%%%%%%
\subsection{Correction to the adiabatic response}
%%%%%%%%%%%%%%%%%%%%%%%%%%%%%%%%%%%%%%%%%%%%%%%%%%%
The response in second order  in frequency  
\begin{eqnarray}\label{eq_current_second}
I^{(2)}(t) = -\frac{e^2h}{2}
\int dE\left(-f'(E)\right)
\frac{\partial}{\partial t}\left[\nu_2^2\frac{\partial
  U_2(t)}{\partial t}\right.
\nonumber \\
\left.+\frac{\partial
  U_1(t)}{\partial t}
 \left[ \nu_1^2
  +2\nu_1\nu_2
 +2\nu_1\nu_d\right] \right]\ ,
\end{eqnarray} 
contains mixed terms in the densities of states of the cavities and the
connecting channel, $\nu_d=\nu_d(E)$, as well. Comprising information about the
entire system, 
it can lead to non-vanishing contributions in the
regime where the adiabatic current response vanishes.  It is interesting
to consider corrections in higher order $\Omega$,
Eq.~(\ref{eq_current_second}), which when $\langle(I^{(1)})^2\rangle$ vanishes
are dominant. Independently of the temperature regime, the correction to 
$\langle I^2\rangle$ in second order in $\Omega$, for the parameters where
$\langle(I^{(1)})^2\rangle$ in first order in $\Omega$ vanishes is always
zero. The leading term in $\Omega$ of $\langle I^2\rangle$ is then at least of
third power in $\Omega$. \par
%\subsubsection{Zero Temperature}
 At zero temperature and small QPC transmission under the conditions given in
 Eqs.~(\ref{eq_conditions})  
we find that $\langle I^2\rangle\sim \langle \left(I^{(2)}\right)^2 \rangle $
is of the order  
$\langle \left(I^{(2)}\right)^2 \rangle \sim (e^2/w)
\left(\tau_1/Tw\right)^{2}$, where $ \tau_{1}/T$ is the dwell time for an
electron in the first cavity. In comparison, the
contribution
%~\cite{noteadiabatic}
 stemming from the first order in  
frequency current away from resonance is $\langle
\left(I^{(1)}\right)^2\rangle \sim e^2/w$, see
Eq.~(\ref{eq_breit_wigner}). We find that in the non-linear regime the adiabaticity condition is  $I^{(2)}/I^{(1)} \sim \Omega\tau/T^2\ll 1$, which differs strongly from the one in linear response,  $\Omega\tau/T \ll 1$.\par    
%\subsubsection{High Temperatures, $k_\mathrm{B}T\gg\Delta$}
At high temperature and at the parameters nullifying
Eq.~(\ref{eq_highT_first}), 
$\langle \left(I^{(2)}\right)^2 \rangle$ is in general not zero and it
depends additionally on the transmission of the QPCs and the inverse of the
density of states of the connecting edge state, $\Delta_d$. 
However, in contrast to the
low-temperature regime $\langle(I^{(2)})^2\rangle$ can be nullified by
introducing further transmission-dependent conditions. These conditions are directly obtained from Eq. (\ref{eq_current_second}) and read
 for cavities of unequal lengths,
$\frac{2-T_1}{2T_1}+\frac{\Delta_1}{\Delta_d}-\frac{3T_2-2}{2T_2}
\frac{\Delta_1}{\Delta_2}=0$.
%\par
%%%%%%%%%%%%%%%%%%%%%%%%%%%%%%%%%%%%%%%%%%%%%%%%%%
\subsection{Nonadiabatic step-potential modulation}
%%%%%%%%%%%%%%%%%%%%%%%%%%%%%%%%%%%%%%%%%%%%%%%%%%
In an experimental set-up, instead of by a sinusoidal modulation, the system
is often driven by a square-pulse potential, where a treatment in the highly
nonadiabatic regime is required. For the following analysis we start from 
Eqs.~(\ref{eq_full_s}) and~(\ref{eq_current_general}), and limit ourselves to
the high-temperature regime. We are interested in a step
potential which is the limit of a periodic square-pulse modulation with
infinitely long period. In principle,
the potential at the two cavities can have different amplitudes and can be 
switched on at different times $t_1^0$ and 
$t_2^0$, with
$t_2^0=t_1^0+\tau_d+\Delta t^0$,
the sum of 
the switching time of the
first cavity $t_1^0$, the time a particle needs to pass through the connecting
edge state $\tau_d$ and a time-delay $\Delta t^0$,
where here we choose $\Delta t^0=0$. The step potentials at the two cavities
read $U_j(t)=U_j$ if $ t \geq t_j^0$ and $0$ otherwise.
The cavities' response to the potentials decays with a characteristic time
given by the bigger value of
$\{\tau_1/\mathrm{ln}(1/R_1),\tau_2/\mathrm{ln}(1/R_2)\}$. After a waiting
time which is much bigger than the decay time, the charge emitted by the
system equals the sum of the charges that would be emitted by two completely
independent cavities and is given by
$Q=e\frac{eU_1}{\Delta_1}+e\frac{eU_2}{\Delta_2}$.
While this charge is nullified for
$\frac{eU_1}{\Delta_1}=-\frac{eU_2}{\Delta_2}$, the nullifying of the
integral of the squared current can in general not be reached, meaning that an
AC current is generated. \par
To find some simple analytical results, let us
restrict ourselves to the limit of identical ($r_1=r_2$ and $L_1=L_2$), weakly
coupled cavities ($T_1=T_2=T\rightarrow 0$) and consider the interesting case
where the total charge is nullified and $\langle I^2\rangle$ is suppressed,
i.e. $U_1=-U_2\equiv U$. 
For a single cavity as well as for the double-cavity system we find the
time-integral over the squared current to be of the form
\begin{equation}
\langle I^2\rangle  
= \frac{e^2}{h}\frac{\left(eU\right)^2}{\Delta}\ F(T,U)\ .
\end{equation}
The function $F$ for a system with a single cavity is given by
$F_\mathrm{single}=T/2$. For the double-cavity system with equal lengths,
$F(T,U)$
oscillates in the potential difference  with a phase
$\phi=2\pi e(U_1-U_2)/\Delta=4\pi eU/\Delta $.
We find 
$F_\mathrm{double}=T^3/[2-2\cos(\phi)]$, if $\phi\neq2n\pi$,
$F_\mathrm{double}=T^{3}/2$ if $\phi=(2n+1)\pi$ and $F_\mathrm{double}=T/4$,
if $\phi=2n\pi$.  
The time integral of the squared
current is
of the same order for the system of a single and a double cavity, showing that
the coupling between the two cavities 
is important
in the highly
nonadiabatic regime. This
is indicated already in
Eq.~(\ref{eq_current_second}), where in second order $\Omega$ mixed terms in
the densities of states of the two cavities appear.
%%%%%%%%%%%%%%%%%%%%%%%%%%%%%%%%%%%%%%%%%%%%%%%%%%
\section{Conclusions}
We investigated the AC current response of a two-particle emitter consisting
of a double-cavity system, and propose it as an efficient tool for counting
electrons emitted at high speed. The square of the total current integrated
over one period shows a pronounced dip when the two cavities are
synchronized. We extract the conditions for  perfect counting by complete
current nullification and show that in the adiabatic regime the counting
efficiency is maintained at small deviations from the obtained
conditions. In the highly nonadiabatic regime, current nullification can in general not be obtained. However, in principle, pulsed cavities can be used to analyze single events. 
%%%%%%%%%%%%%%%%%%%%%%%%%%%%%%%%%%%%%%%%%%%%%%%%%%
\acknowledgments
J.~S. and M.~B. acknowledge the support of the Swiss NSF, the National Center of Competence in
Research MANEP and the European STREP
project SUBTLE.


\begin{thebibliography}{24}

\bibitem{buttiker93} M. B\"uttiker, H. Thomas, and A. Pr\^etre, 
                     Phys. Lett. A {\bf 180}, 364 (1993). 
                     

\bibitem{gabelli06}  J. Gabelli, %\textit{et al.},
                     G. F\`eve, J.-M. Berroir, B. Pla\c{c}ais,
                     A. Cavanna, B. Etienne, Y. Jin, and D. C. Glattli, 
                     Science {\bf 313}, 499 (2006).
              
%\bibitem{rq}        
\bibitem{nigg06} S. E. Nigg, R. Lopez, and M. B\"uttiker, 
                     Phys. Rev. Lett. {\bf 97}, 206804 (2006).

\bibitem{wang07}                     J.	 Wang, B. Wang, and H. Guo,
                     Phys Rev. B {\bf 75} 155336 (2007).  

\bibitem{nigg08}                     S. E. Nigg and M. B\"uttiker, 
                     Phys. Rev. B {\bf 77}, 085312 (2008).                 

\bibitem{feve07}  
                     G. F\`eve, % \textit{et al.},
                     A. Mah\'e,  J.-M. Berroir, T. Kontos,
                     B. Pla\c{c}ais,  D. C. Glattli, A. Cavanna, B. Etienne, 
                     and  Y. Jin, 
                     Science {\bf 316}, 1169 (2007). 
              
              
\bibitem{moska08} 
                     M. Moskalets, P. Samuelsson, and M. B\"uttiker,
                     Phys. Rev. Lett. {\bf 100}, 086601 (2008). 
        
\bibitem{keeling08}                     J. Keeling, A. V. Shytov, and L. S. Levitov, 
                     (unpublished),  arXiv:0804.4281. 
         
\bibitem{olkho08}                     S. Ol'khovskaya, J. Splettstoesser, M. Moskalets, and M. B\"uttiker, 
                     Phys. Rev. Lett. {\bf 101},  166802 (2008).
       
         
%\bibitem{pumps}      

\bibitem{blumenthal07} M. D. Blumenthal, %\textit{et al.},
                     B. Kaestner, L. Li, S. Giblin, T. J. B. M. Janssen, 
                     M. Pepper, D. Anderson, G. Jones, and D. A. Ritchie,
                     Nature Physics {\bf 3}, 343 (2007).

\bibitem{kaestner08}		   B. Kaestner, %\textit{et al.},
		   V. Kashcheyevs, S. Amakawa, M. D. Blumenthal, L. Li,
                      T. J. B. M. Janssen, G. Hein, K. Pierz, T. Weimann,
                      U. Siegner, and H. W. Schumacher, 
		    Phys. Rev. B {\bf 77}, 153301 (2008).

\bibitem{fujiwara08}		     %A. Fujiwara, \textit{et al.},
		     A. Fujiwara,
		     K. Nishiguchi,  and Y. Ono,
		     Appl. Phys. Lett. {\bf 92}, 042102 (2008).

\bibitem{maire08}		     N. Maire, % \textit{et al.},
		     F. Hohls, B. Kaestner, K. Pierz, H. W. Schumacher,
                        R. J. Haug, 
		     Appl. Phys. Lett. {\bf 92}, 082112 (2008).

\bibitem{kaestner08_2}		     B. Kaestner, %\textit{et al.},
		     V. Kashcheyevs, G. Hein, K. Pierz, U. Siegner,
		      H. W. Schumacher,
		     Appl. Phys. Lett. {\bf 92}, 192106 (2008).

              

\bibitem{lu2003}     W. Lu, % \textit{et al.},
                     Z. Ji, L. Pfeiffer, K. W. West, A. J. Rimberg,
                     Nature {\bf 423}, 422 (2003).

\bibitem{fujisawa2004} T. Fujisawa, %\textit{et al.},
                       T. Hayashi, Y. Hirayama, H. D. Cheong, and Y. H. Jeong,
                       Appl. Phys. Lett. {\bf 84}, 2343 (2004).

\bibitem{bylander2005} J. Bylander, T. Duty, and P. Delsing,
                       Nature {\bf 434}, 361 (2005).
  
\bibitem{gustavsson06} S. Gustavsson, %\textit{et al.},
                       R. Leturcq, B. Simovic, R. Schleser, T. Ihn,
                       P. Studerus, K. Ensslin, D. C. Driscoll, and A. C. Gossard,
                       Phys. Rev. Lett. {\bf 96}, 076605 (2006).

\bibitem{fujisawa06} T. Fujisawa, %\textit{et al.},
                     T. Hayashi, R. Tomita, and Y. Hirayama,
                     Science {\bf 312}, 1634 (2006).

\bibitem{fricke07} C. Fricke, % \textit{et al.},
                   F. Hohls, W. Wegscheider, and R. J. Haug,
                   Phys. Rev. B {\bf 76}, 155307 (2007).  
                            
\bibitem{reilly}   D. J. Reilly, C. M. Marcus, M. P. Hanson, and A. C. Gossard, Appl. Phys. Lett. {\bf 91},
                   162101 (2007). 
               
               
\bibitem{comment_random}  Counting in the presence of partition noise will be the subject of future                              
                                                   investigations.
             
             
%\bibitem{notenoise} 
 
 \bibitem{fertig87} H. A. Fertig and B. I. Halperin, Phys. Rev B {\bf 36}, 7969
                   (1987). 


% \bibitem{quantum_pump} 
\bibitem{brouwer98} P. W. Brouwer, Phys. Rev. B {\bf 58}, R10135 (1998).

\bibitem{avron00}                        J. E. Avron, %\textit{et al.}, 
                        A. Elgart, G. M. Graf, and L. Sadun,
                        Phys. Rev. B {\bf 62}, R10618 (2000).
                        
\bibitem{vavilov05}                        M. G. Vavilov, L. DiCarlo, and C. M. Marcus, 
                        Phys. Rev. B {\bf 71}, 241309(R) (2005).

\bibitem{arrachea05}                        L. Arrachea, Phys. Rev. B {\bf 72}, 125349 (2005).

\bibitem{splett05}                        J. Splettstoesser, %\textit{et al.}, 
                        M. Governale, J. K\"onig, and R. Fazio,
			Phys. Rev. Lett. {\bf 95}, 246803 (2005).

\bibitem{graf08}			G. M. Graf and G. Ortelli, 
			Phys. Rev. B {\bf 77}, 033304 (2008).  


             
%\bibitem{noteadiabatic}  

\end{thebibliography}
\end{document}